# Detection of E-cyanomethanimine towards Sagittarius B2(N) in the Green Bank Telescope PRIMOS Survey

Daniel P. Zaleski[1], Nathan A. Seifert[1], Amanda L. Steber[1], Matt T. Muckle[1], Ryan A. Loomis[1], Joanna F. Corby[2], Oscar Martinez, Jr.[3], Kyle N. Crabtree[3], Philip R. Jewell[4], Jan M. Hollis[5], Frank J. Lovas[6], David Vasquez[1], Jolie Nyiramahirwe[1], Nicole Sciortino[1], Kennedy Johnson[1], *Michael C. McCarthy[3], *Anthony J. Remijan[4], and *Brooks H. Pate[1]

[1] *Department of Chemistry, University of Virginia, McCormick Rd, Charlottesville, VA. 22904, USA*

[2] *Department of Astronomy, University of Virginia, McCormick Rd, Charlottesville, VA 22904, USA.*

[3] *Harvard-Smithsonian Center for Astrophysics, 60 Garden St., Cambridge, MA 02138, and School of Engineering & Applied Sciences, Harvard University, 29 Oxford St., Cambridge, MA 02138.*

[4] *National Radio Astronomy Observatory, 520 Edgemont Rd., Charlottesville, VA 22903-2475.*

[5] *NASA Goddard Space Flight Center, Greenbelt, MD, 20771*

[6] *National Institute of Standards and Technology, Gaithersburg, MD 20899*

*corresponding author, email address: mccarthy@cfa.harvard.edu (Michael C. McCarthy)
*corresponding author, email address: aremijan@nrao.edu (Anthony J. Remijan)
*corresponding author, email address: bp2k@virginia.edu (Brooks H. Pate)

**Abstract**

The detection E-cyanomethanimine (E-HNCHCN) towards Sagittarius B2(N) is made by comparing the publicly available Green Bank Telescope (GBT) PRIMOS survey spectra (Hollis et al.) to laboratory rotational spectra from a reaction product screening experiment. The experiment uses broadband molecular rotational spectroscopy to monitor the reaction products produced in an electric discharge source using a gas mixture of $NH_3$ and $CH_3CN$. Several transition frequency coincidences between the reaction product screening spectra and previously unassigned interstellar rotational transitions in the PRIMOS survey have been assigned to E-cyanomethanimine. A total of 8 molecular rotational transitions of this molecule between 9 and 50 GHz are observed with the GBT. E-cyanomethanimine, often called the HCN dimer, is an important molecule in prebiotic chemistry because it is a chemical intermediate in proposed synthetic routes of adenine, one of the two purine nucleobases found in DNA and RNA. New analyses of the rotational spectra of both E-cyanomethanimine and Z-cyanomethanimine that incorporate previous mm-wave measurements are also reported.

**Section 1.    INTRODUCTION**

One important goal of the field of astrobiology is the identification of chemical synthesis routes for the production of molecules important in the development of life that are consistent with the chemical inventory and physical conditions on newly formed planets. One mechanism for seeding planets with chemical precursors is delivery by outer solar system bodies, like comets or meteorites (Chyba 1990). These objects can be chemical reservoirs for the molecules produced in the interstellar medium during star and planet formation. The chemical inventory of these objects includes the molecules that are directly incorporated from the interstellar medium and molecules subsequently formed by chemical processing of the interstellar species (Chyba et al. 1990, Chyba & Sagan 1992, Oró et al. 1992). This subsequent chemical processing can synthesize larger, more complex molecules that are more directly relevant to prebiotic chemistry from the simpler molecules that can be formed in the interstellar medium. The identification of molecules in the interstellar medium is a key step in understanding the chemical evolution from simple molecular species to molecules of biological relevance and radio astronomy has played the dominant role in identifying the chemical inventory of the interstellar medium (Herbst & van Dishoeck 2009).

The star forming region Sagittarius(Sgr) B2(N) is the richest interstellar chemical environment currently known. In particular, many of the detections of complex organic molecules by radio astronomy have been made towards this source (see e.g. Belloche et al. 2008, Belloche et al. 2009, Friedel et al. 2004, Nummelin et al. 1998, and references therein). In fact, roughly half of the 170 molecules that have been detected in space were first detected toward the Sgr B2(N) region. One puzzling result is that many of the large molecules identified in this source are found to have cold rotational temperatures (~ 10 K) despite the fact that most models of complex molecule formation require high-temperature (100-200 K) processing of interstellar ices. The fact that many of the large interstellar organic molecules in Sgr B2(N) are rotationally cold has made observations using the Green Bank Telescope (GBT) at microwave frequencies, which are a good match to the peak intensity distribution of the rotational spectrum at low temperature, highly successful for molecule identification in this rich source.

The detection of interstellar E-cyanomethanimine (E-HNCHCN), with the molecular structure shown in Figure 1, towards Sgr B2(N) in a publicly available GBT spectral survey (Hollis et al.) has been made using a reaction product screening method. This method is similar in concept to the well-known Urey-Miller spark-initiated chemistry experiments (Miller 1953, Miller & Urey 1959). The idea is to identify all chemical products in the reaction of high-abundance interstellar molecules in an electric discharge. The reaction chemistry is performed in a pulsed electric discharge molecular beam source (McCarthy et al. 2000). The rotational spectra of product molecules are measured using broadband molecular rotational spectroscopy in a chirped-pulse Fourier transform microwave spectrometer. Direct comparison of the laboratory spectrum from the complex mixture of reaction products to the interstellar survey spectrum provides an efficient way to focus the molecular identification process on molecules of high interest because their rotational transitions overlap with unassigned interstellar transitions. Based on the predicted reaction chemistry, candidate product molecules can then be identified using available catalogs (like Splatalogue.net), or a combination of the acquired spectroscopic data and quantum chemistry predictions of the relevant spectral parameters.

**Section 2.    OBSERVATIONS AND RESULTS**

*Section 2.1.    Astronomical Observations*

E-cyanomethanimine transitions were initially identified in the survey spectra from the National Radio Astronomy Observatory's (NRAO) 100m Robert C. Byrd Green Bank Telescope (GBT) PRebiotic Interstellar MOlecule Survey (PRIMOS) Legacy Project conducted from 2008 January through 2011 July. Additional observations targeted at E-cyanomethanimine frequencies not covered in this survey were completed on 2012 October 11. These observations targeted the $3_{13}$-$2_{12}$, $3_{03}$-$2_{02}$, and the $3_{12}$-$2_{11}$ rotational transitions at 28158.894 MHz, 28709.707 MHz, and 29273.956 MHz, respectively. Observations were made in the OFF-ON position-switching mode, with the OFF position 1°; east in azimuth with respect to the ON-source position, the Sgr B2(N) Large Molecule Heimat (LMH) ($\alpha$J2000 = 17h47m19s.8, $\delta$J2000 = -28°22'17.0"). Each OFF–ON observing cycle consisted of 2 minutes in the OFF-source position followed by 2 minutes in the ON-source position. Automatically updated dynamic pointing and focusing corrections were used based on real-time temperature measurements of the structure input to a thermal model of the GBT; zero points were typically adjusted every 2 hours or less using the quasar 1733-130 for calibration (see e.g. Remijan et al. 2008). Antenna temperatures were recorded on the Ta* scale (Ulich & Haas 1976) with estimated absolute flux calibration uncertainties of up to 20% in some observing bands. The data reduction was completed using the GBTIDL reduction software, with both polarizations (when available) averaged to improve signal-to-noise in the final data reduction. All data were continuum subtracted to allow for Gaussian profile fitting of each detected spectral feature.

*Section 2.2.    Laboratory Measurements*

Broadband molecular rotational spectra of the product molecules produced in a pulsed supersonic expansion of acetonitrile ($CH_3CN$) and ammonia ($NH_3$) through a high voltage (2 kV) DC discharge (McCarthy et al. 2000) were acquired using a chirped-pulse Fourier transform microwave (CP-FTMW) spectrometer at the University of Virginia. The reaction product screening measurement was performed in two different frequency ranges, 6.5-18.5 GHz (Brown et al. 2008) and 25-40 GHz (Zaleski et al. 2012) with $2 \times 10^5$ and $3.5 \times 10^5$ signal averages, respectively.

Due to the presence of two nitrogen nuclei that produce nuclear quadrupole hyperfine structure, it was necessary to perform higher spectral resolution pulsed molecular beam measurements using the coaxial-orientation cavity Fourier transform microwave spectrometer at the Harvard-Smithsonian Center for Astrophysics (Grabow et al. 2005). Measurements of the rotational spectra of both E-cyanomethanimine and Z-cyanomethanimine were performed. Additional laboratory measurements of rotational transitions in the 40-50 GHz frequency range were performed with a chirped-pulse Fourier transform microwave spectrometer (Zaleski 2012). In total, the frequency range between 9 and 50 GHz was covered. These new laboratory measurements have been combined with previous microwave and mm-wave measurements (Takano et al. 1990) in a global fit and the results from an analysis using the Watson S reduction (Watson 1977) are presented in Table 1. The transition frequencies used for the global analysis, including the new measurements from this work, are available online at www.splatalogue.net. Spectral analysis was performed using SPFIT/SPCAT (Pickett 1991). The rotational spectroscopy parameters for E-cyanomethanimine and Z-cyanomethanimine are compared to calculated values obtained using the Gaussian09 software package (Frisch et al. 2009) at the MP2/6-311++G(d,p) level of theory in Table 1.

*Section 2.3.    Interstellar Identification of E-cyanomethanimine*

The identification of overlapping transitions of E-cyanomethanimine from the laboratory reaction product screening experiment and the GBT PRIMOS survey spectrum are shown in Figure 2a where the laboratory data (green trace) is overlaid directly on the GBT PRIMOS spectra (black trace). In this comparison, the frequencies of the PRIMOS survey spectra have been shifted to the laboratory rest frame using the LSR source velocity of +64 km/s that is commonly found in Sgr B2(N). An important feature of this transition overlap is that the nuclear quadrupole hyperfine structure of this rotational transition ($1_{01}$-$0_{00}$) is observed in the PRIMOS spectrum as well. This coincidence in transition frequency (related to the principal moments of inertia) and hyperfine structure (determined by the electronic properties of the chemical bonding) is a strong indication that E-cyanomethanimine is the molecule generating the interstellar spectrum. Subsequently a total of 8 molecular transitions of E-cyanomethanimine were detected in the GBT spectrum, out of 9 possible transitions with appreciable intensity in the frequency range of 9-50 GHz. The remaining transition, the $2_{12}$-$1_{11}$, coincides with a recombination line.

The full set of detected interstellar transitions is shown in Figure 2. All transition rest frequencies shown in Figure 2 have been shifted to the nominal LSR source velocity of +64 km/s (blue dashed line). A weaker rotational transition with a velocity component of +82 km/s (red dashed line) is also observed and this velocity component has been found in several other molecules observed with the GBT (see e.g. McGuire et al. 2012, Remijan et al. 2008 and references therein). It can also be seen that all transitions exhibit similar linewidths. Table 2 lists the experimentally observed frequencies (MHz), transition line strength ($D^2$), lower and upper state energies (K), measured continuum temperatures (K) and measured line intensities (K) and linewidths (km/s) from the GBT PRIMOS survey.

The excitation temperature and total beam-averaged column density for E-cyanomethanimine was determined using the formalism described in Remijan et al. (2005). The total beam-averaged column density for a single-element radio telescope is given as

$$N_T = 8.5 \times 10^9 \times \frac{Q_r(\Delta T_A^* \Delta V/\eta_B)}{(T_{ex}-T_c/\eta_B)S\mu^2(e^{-E_l/T_{ex}}-e^{-E_u/T_{ex}})} \quad (1)$$

for spectral line features detected in absorption. In this equation, the line shape is assumed to be Gaussian and we assume that absorption components fill the telescope beam; $\eta_B$ is the telescope beam efficiency as determined by the Ruze formulation (Ruze 1966) and listed for each observing frequency for the GBT in Hollis et al. (2006); $T_{ex}$ is the calculated excitation temperature; $\Delta T_A^*\Delta V$ is the product of the fitted line intensity (mK) and line width (km/s); $Q_r$ is the rotational partition function ($Q_r = 4.4 \times T^{3/2}$); $S\mu^2$ is the product of the transition line strength and the square of the dipole moment (Debye$^2$). $E_u$ is the upper state rotational energy level (K); $T_c$ is the source continuum temperature (K) measured by the GBT from channels free from line emission, and $E_l$ is the lower state rotational energy level (K). It is assumed that $T_c$ measured by the GBT is approximately equal to the local continuum temperature of the E-cyanomethanimine containing regions.

The measured spectral line parameters given in Table 2 were fit to Equation 1 using a linear least squares method and a best fit temperature of 8(2) K and column density of ~1.5(2) x $10^{13}$ cm$^{-2}$ were determined. Other nitriles and nitrile derivatives in Sgr B2(N) have similar cold rotational temperatures, which could be indicative of sub-thermal behavior (Loomis et al. 2012, Remijan et al. 2005). Additionally, this noted commonality in the rotational temperatures of these species suggests they may be localized within the same environments in Sgr B2(N), a source with substantial structure on spatial scales smaller than the GBT beam. The temperature reported here is not necessarily indicative of the formation temperature, either in the gas phase or desorbed from a grain surface, that takes place in hot molecular cores.

## Section 3. DISCUSSION

The reaction product screening measurements using the CH$_3$CN and NH$_3$ discharge chemistry produce two additional isomers of E-cyanomethanimine, shown in Figure 1. The cis-trans isomer (about the C=N bond), Z-cyanomethanimine, is produced in about equal abundance. This laboratory result is expected based on the isomerization potential calculated from quantum chemistry. The Z- form is calculated by *ab initio* to be about 370 K more stable (in agreement with the experimental energy difference of 309(72) K reported by Takano et al. 1990). Furthermore, there is a high barrier to isomerization (15.95 kK) between the E- and Z-isomers. The production of cyanomethanimine in the laboratory is believed to occur by the exothermic recombination of the primary radicals generated from the two reagents (·CH$_2$CN and ·NH$_2$) followed by dehydrogenation (shown in Figure 3). Prior to the supersonic expansion, the reactants are generated in a high temperature environment where they acquire sufficient energy imparted by the electric discharge to undergo chemical processes.

$$\cdot NH_2 + \cdot CH_2CN \rightarrow NH_2CH_2CN \quad (2) \quad (E_a = 0 \text{ K}, \Delta E = -39.29 \text{ kK})$$
$$NH_2CH_2CN \rightarrow \text{Z-HNCHCN} + H_2 \quad (3) \quad (E_a = 58.05 \text{ kK}, \Delta E = 16.56 \text{ kK})$$
$$NH_2CH_2CN \rightarrow \text{E-HNCHCN} + H_2. \quad (4) \quad (E_a = 58.89 \text{ kK}, \Delta E = 16.93 \text{ kK})$$

When formed in a pulsed molecular beam, both product molecules are often observed due the rapid cooling environment of the pulsed adiabatic expansion that traps the populations behind the large isomerization barrier. A similar reaction process occurs for ethanimine, a molecule where both E- and Z-isomers are observed in the laboratory and in Sgr B2(N) (Loomis et al. 2012).

Despite being produced in near equal abundance, the intensities of the rotational transitions from Z-cyanomethanimine are significantly weaker because of the difference in the dipole moments. The $\mu_a$ and $\mu_b$ dipole components for the E- form are 10.87(7)x10$^{-30}$ Cm and 8.37(7)x10$^{-30}$ Cm [3.25(2) D and 2.51(2) D], respectively compared to $\mu_a$=4.50(33)x10$^{-30}$ Cm and $\mu_b$= 1.3(17)x10$^{-30}$ Cm [$\mu_a$=1.35(10) D and $\mu_b$=0.4(5) D] for the Z- form (Takano et al. 1990). Uncertainties here are Type A, coverage factor k=1 (1 sigma) (Taylor & Kuyatt 1994). As a result, the a-type rotational transitions of Z-cyanomethanimine are intrinsically 6 times weaker than those of E-cyanomethanimine. Given the current sensitivity limits (~2-5mK RMS) of the GBT PRIMOS survey, the nondetection of the less polar Z-isomer is perhaps not surprising.

A third isomer of cyanomethanimine was also detected in the laboratory, N-cyanomethanimine (see Figure 1). N-cyanomethanimine structurally differs from E- and Z-cyanomethanimine by the position of the nitrile group on the nitrogen atom (see Figure 1), and its rotational spectrum has been previously measured (Bak et al. 1978, Bak et al. 1979, Winnewisser & Winnewisser 1984). There are weak, unassigned transitions in the PRIMOS survey spectrum that coincide with the $2_{02}$-$1_{01}$ (20915.29 MHz) and $3_{03}$-$2_{02}$ (31366.52 MHz) rotational transitions of N-cyanomethanimine, but there is currently insufficient data to make a definitive detection.

The detection of E-cyanomethanimine in Sgr B2(N) poses a challenge to the current understanding of mechanistic interstellar chemistry. The formation of cyanomethanimine by gas-phase synthetic routes that are compatible with low-temperature interstellar environments was studied by Smith et al. 2001, and it was concluded that the mechanisms investigated were incapable of producing significant amounts of the molecule. A recent mechanistic study extended this work to examine possible acid-catalyzed reaction pathways that become available when molecules are protonated under interstellar conditions (Yim & Choe 2012). This quantum chemistry study also concluded that there was no viable gas-phase route to the interstellar production of cyanomethanimine. Much of the recent effort to understand the production of complex organic molecules under interstellar conditions has focused on the processing of interstellar ices by high-energy excitation such as cosmic ray bombardment or ultraviolet radiation. We note that the laboratory production mechanism likely involves radical recombination chemistry – although through reactions that are initiated in a high-temperature environment prior to cooling in the pulsed supersonic jet expansion. Future experimental or theoretical work of interstellar ice chemistry may be able to provide a suitable mechanistic pathway for cyanomethanimine production.

Cyanomethanimine is important to prebiotic chemistry and the formation of adenine ($C_5H_5N_5$), one of the two purine nucleobases that are critical in the RNA-world scenario for formation of life on Earth. In addition to its role in the nuclei acids DNA and RNA, adenine plays a key role in several biological chemistry processes including cellular respiration, in the form of the energy-rich adenosine triphosphate (ATP), and protein synthesis (Carey 1992). One proposed prebiotic synthetic route for the formation of adenine is a sequence of condensation or oligomerization reactions of HCN (Sanchez et al. 1966, Schwartz et al. 1982, Borquez et al. 2005). In this scheme, cyanomethanimine is the HCN "dimer" and adenine is the HCN "pentamer" (Oró 1961, Ferris et al. 1966, Ferris et al. 1978). The initial formation of the "HCN dimer" is particularly important because quantum chemistry studies of the reaction pathways suggest that HCN additions subsequent to the formation of the dimer are exothermic (Roy et al. 2007 and references therein). The detection of cyanomethanimine towards Sgr B2(N) shows that it can be formed in the interstellar medium, perhaps by a radical chemistry formation route that could occur in interstellar ices, with the potential for later delivery to a newly formed planet where it can facilitate the production of adenine under prebiotic chemistry conditions.

**Section 4.     CONCLUSIONS**

We have presented the interstellar detection of E-cyanomethanimine in the GBT PRIMOS observations of Sgr B2(N) utilizing astronomical data in public archives and laboratory reaction screening measurements. A column density for E-cyanomethanimine has been determined to be approximately $1.5(2) \times 10^{13}$ cm$^{-2}$. Additionally, a rotational temperature for E-cyanomethanimine has been fit to 8(2) K. Based on frequency coincident overlaps between broadband laboratory spectra and broadband survey spectra, which initially led to the detection of E-cyanomethanimine, there is also tentative evidence for interstellar N-cyanomethanimine. Presently, however, there is not enough data to confirm its presence towards Sgr B2(N). Finally, this work highlights the challenge posed by the detection of E-cyanomethanimine to astronomical chemical models that suggest it should not be formed under interstellar gas-phase conditions.

**ACKNOWLEDGEMENTS**

This work was supported by the NSF Centers for Chemical Innovation (CHE-0847919) and NSF Chemistry (CHE-1213200). Additional support was provided by the Virginia-North Carolina Alliance, a NSF Louis Stokes Alliance for Minority Participation (HRD-1202181) and the National Radio Astronomy Observatory. The National Radio Astronomy Observatory is a facility of the National Science Foundation operated under cooperative agreement by Associated Universities, Inc. Finally, we thank the anonymous referee for a favorable review and valuable comments.

**Table 1**
**Spectral Parameters for E-cyanomethanimine and Z-cyanomethanimine**

|  | E-cyanomethanimine | | Z-cyanomethanimine | |
|---|---|---|---|---|
|  | Experiment[a] | MP2/6-311++G(d,p)[1] | Experiment[a] | MP2/6-311++G(d,p)[1] |
| A (MHz) | 62695.094(24) | 61297.662 | 54173.1(50) | 53903.7 |
| B (MHz) | 4972.04643(81) | 4918.70720 | 5073.86506(86) | 4996.28000 |
| C (MHz) | 4600.29460(89) | 4553.33410 | 4632.39090(74) | 4572.46290 |
| DJ (kHz) | 1.8704(55) | 1.8079 | 2.4737(74) | 2.3088 |
| DJK (kHz) | -104.55(17) | -103.15 | -103.31(21) | -103.72 |
| d1 (kHz) | -0.3272(57) | -0.3269 | -0.4961(82) | -0.4551 |
| d2 (kHz) | 0.0221(71) | 0.0167 | -0.0321(55) | -0.0250 |
| CN 1.5Xaa (MHz) | -6.192(10) | -5.331 | -6.018(32) | -5.185 |
| CN .25(Xbb-Xcc) (MHz) | -0.1972(57) | -0.1368 | -0.2146(86) | -0.1614 |
| NH 1.5Xaa (MHz) | 1.129(16) | 1.356 | -6.404(31) | -6.142 |
| NH .25(Xbb-Xcc) (MHz) | -2.0642(89) | -1.9661 | -0.8201(82) | -0.7014 |
| N lines | 83 | | 87 | |
| RMS (kHz) | 22.7 | | 19.8 | |

**Notes.** A comparison of experimentally determined spectral parameters (rotational constants, distortion constants, and nuclear hyperfine constants) determined from the Watson S reduction compared to theory. The (global) fits reported here include measured transitions from Takano et al. 1990.

[a] Uncertainties are Type A, coverage factor k=1 (1 sigma) (Taylor & Kuyatt 1994).

**References**. (1) Frisch et al. 2009.

**Table 2**
**Observed Interstellar Transitions of E-Cyanomethanimine**

| Transition | | | Frequency[a] | | | | | + 64 km s$^{-1}$ | | + 82 km s$^{-1}$ | |
|---|---|---|---|---|---|---|---|---|---|---|---|
| $J_{K_aK_c}' - J_{K_aK_c}''$ | F' - F'' | $I_{12}' - I_{12}''$ | (MHz) | $E_l$ (K) | $E_u$ (K) | $S_{ij}\mu^2$ (D$^2$) | $T_c$ (K) | $\Delta T^b$ (mK) | $\Delta V^b$ (km s$^{-1}$) | $\Delta T^b$ (mK) | $\Delta V^b$ (km s$^{-1}$) |
| $1_{01} - 0_{00}$ | 2 - 1 | 2 - 1 | 9571.312(5) | 0 | 0.4590 | 12.0409 | 24.72 | -22.6(1.1) | 15.2(0.9) | - | - |
| $1_{01} - 0_{00}$ | 3 - 2 | 2 - 2 | 9572.505(5) | 0 | 0.4590 | 24.6016 | 24.72 | -27.4(1.4) | 22.1(1.3) | - | - |
| $2_{02} - 1_{01}$ | 4 - 3 | 2 - 2 | 19142.905(5) | 0.4594 | 1.3788 | 38.0689 | 9.22 | -38.0(0.9) | 11.3(0.3) | -22.9(1.0) | 8.8(0.4) |
| $2_{11} - 1_{10}$ | 4 - 3 | 2 - 2 | 19516.889(5) | 3.2506 | 4.1873 | 28.5156 | 8.66 | -9.1(0.9) | 6.2(0.6) | -6.0(1.1) | 4.2(0.9) |
| $3_{13} - 2_{12}$ | 5 - 4 | 2 - 2 | 28158.894(5) | 4.1340 | 5.4850 | 44.2225 | 13.43 | -39.7(2.2) | 18.8(1.1) | - | - |
| $3_{03} - 2_{02}$ | 5 - 4 | 2 - 2 | 28709.707(5) | 1.3781 | 2.7565 | 49.8436 | 10.41 | -100.7(2.4) | 10.4(0.3) | -34.1(2.4) | 11.3(0.9) |
| $3_{12} - 2_{11}$ | 5 - 4 | 2 - 2 | 29273.956(5) | 4.1873 | 5.5917 | 44.2225 | 10.05 | -33.3(2.9) | 7.6(0.7) | - | - |
| $4_{04} - 3_{03}$ | 6 - 5 | 2 - 2 | 38271.019(5) | 2.7561 | 4.5935 | 60.9961 | 2.62 | -42.7(1.3) | 8.7(0.3) | -22.2(1.1) | 12.2(0.6) |
| $5_{05} - 4_{04}$ | 7 - 6 | 2 - 2 | 47824.964(40) | 4.5928 | 6.8878 | 72.0801 | 3.17 | -25.7(2.8) | 6.4(0.7) | - | - |

**Notes.** Lists of the experimentally observed frequencies, intensities, and linewidths in the GBT PRIMOS survey. Because the linewidths of the PRIMOS survey are unable to resolve the hyperfine structure resulting from two quadrupolar nuclei (other than the $1_{01}$-$0_{00}$ where two components are reported), only the dominant component is reported in the table above. The nitrogen spins are coupled to an intermediate spin $I_{12}$, where $I = I_1 + I_2$. The intermediate spin is then coupled to the rotational angular momentum, $F = J + I_{12}$. The subscript 1 refers to the CN nitrogen, and the subscript 2 refers to the NH nitrogen. This is the same labeling convention reported in Krause & Sutter 1992. The beamsize varies from 84" at 9 GHz to 16" at 46 GHz. More information on the calculation of the beamsize can be found in Hollis et al. 2006.

[a] Uncertainties are Type B, coverage factor k=1 (1 sigma) (Taylor & Kuyatt 1994).
[b] Uncertainties are Type A, coverage factor k=1 (1 sigma) (Taylor & Kuyatt 1994).

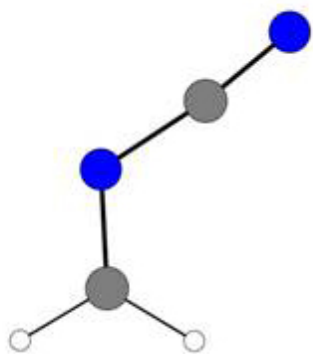

N-cyanomethanimine   E = 3970 K

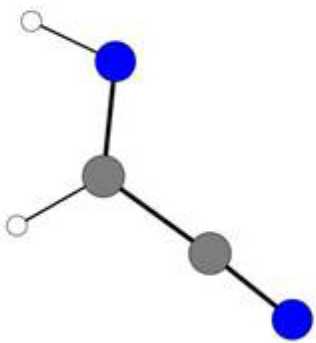

E-cyanomethanimine   E = 370 K

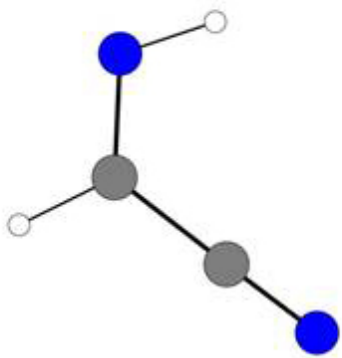

Z-cyanomethanimine   E = 0 K

**Figure 1.** Optimized structures of three isomers of cyanomethanimine. Energies have been calculated at the MP2/6-31G(d,p) level of theory. Blue atoms are nitrogen, grey are carbon, and white are hydrogen.

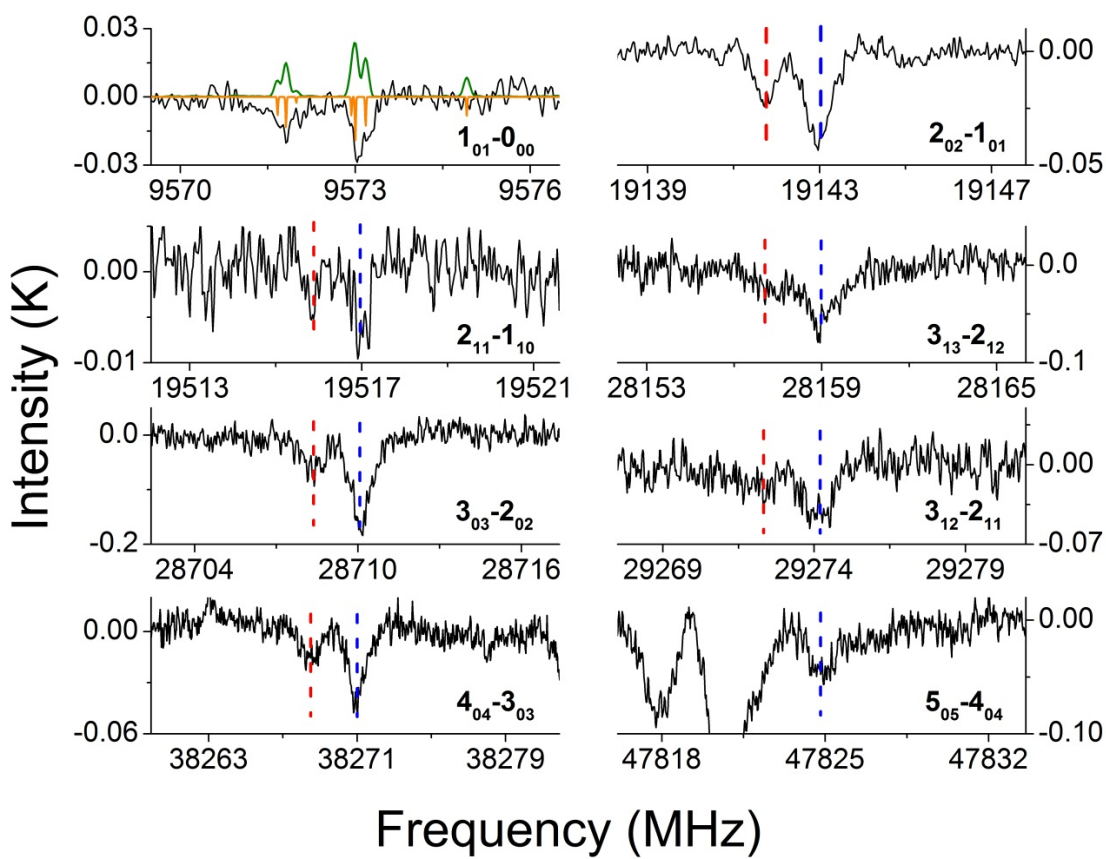

**Figure 2.** Observed transitions of E-cyanomethanimine towards Sgr B2(N) (black trace). Frequencies have been shifted for +64 km/s. The blue drop lines indicate the +64 km/s component, and the red drop lines indicate the +82 km/s component. The strong features near the $5_{05}$-$4_{04}$ transition belong to acetaldehyde, and are likely obscuring the +82 km/s component of E-HNCHCN. The green trace shows the laboratory spectrum, and is only shown for the $1_{01}$-$0_{00}$ transition since it is the only transition with resolvable hyperfine structure in the GBT. The orange trace is an 8 K (rotational temperature) simulation showing the hyperfine structure. Both features in the first panel represent the +64 km/s component.

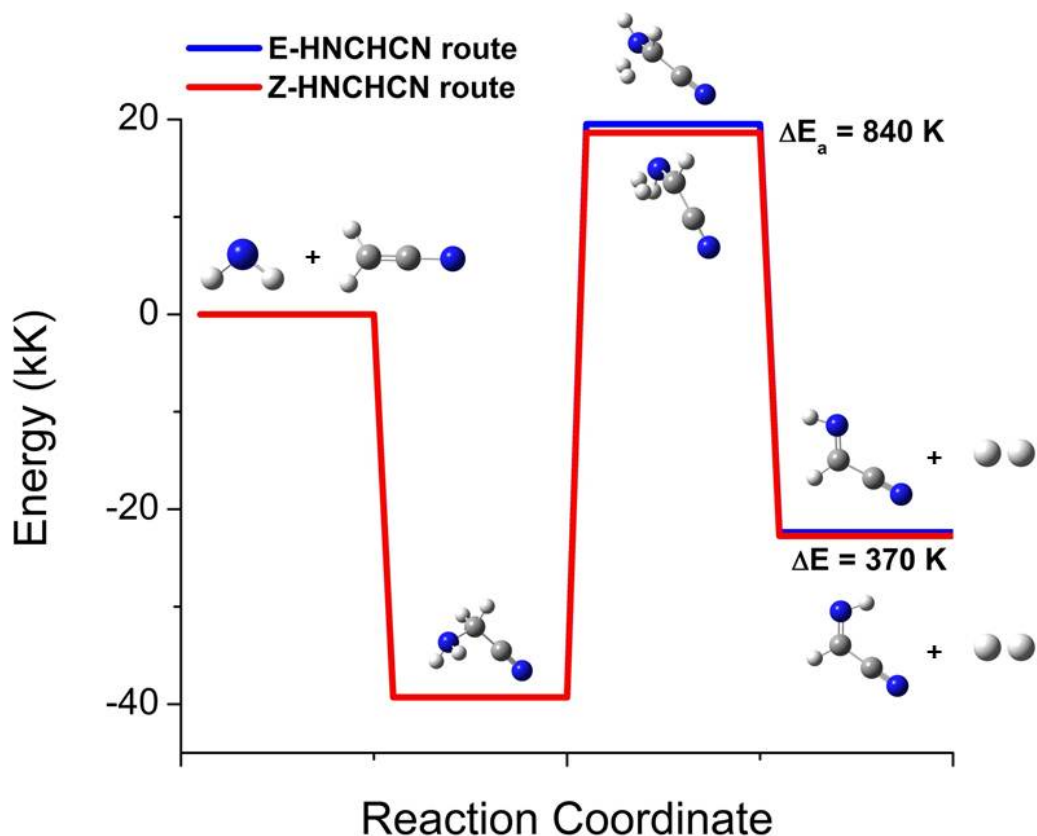

**Figure 3.** Reaction coordinate diagram showing the proposed discharge induced laboratory formation route of E- and Z-cyanomethanimine in the supersonic expansion. The difference in energy between the two different transition states (each leading to a different product) is about 840 K. The difference in energy between the products is 370 K (or the difference between equations 3 and 4). Both products are observed in the laboratory experiment, but only the E-form is observed in the GBT PRIMOS survey, likely due to its larger 3D dipole moment (compared to 1D for the Z-form). The calculations were performed at the MP2/6-31G(d,p) level of theory. Blue atoms are nitrogen, grey are carbon, and white are hydrogen.